\documentclass[8.5pt,twoside,twocolumn]{article}
\usepackage[super,sort&compress,comma]{natbib} 
\usepackage{mhchem}
\usepackage{times,mathptmx}
\usepackage{sectsty}
\usepackage{balance} 

\usepackage{graphicx} %eps figures can be used instead
\usepackage{lastpage}
\usepackage[format=plain,justification=raggedright,singlelinecheck=false,font=small,labelfont=bf,labelsep=space]{caption} 
\usepackage{fancyhdr}
\usepackage{color}
\newcommand{\onlinecite}{\citenum}
\newcommand{\ib}[1]{{\color{black}#1}}
\newcommand{\alt}{\raisebox{-0.3ex}{$\stackrel{<}{\sim}$}}

\begin{document}

\thispagestyle{plain}
\fancypagestyle{plain}{
%arXiv \fancyhead[L]{\includegraphics[height=8pt]{headers/LH}}
%arXiv \fancyhead[C]{\hspace{-1cm}\includegraphics[height=20pt]{headers/CH}}
%arXiv \fancyhead[R]{\includegraphics[height=10pt]{headers/RH}\vspace{-0.2cm}}
\renewcommand{\headrulewidth}{1pt}}
\renewcommand{\thefootnote}{\fnsymbol{footnote}}
\renewcommand\footnoterule{\vspace*{1pt}% 
\hrule width 3.4in height 0.4pt \vspace*{5pt}} 
\setcounter{secnumdepth}{5}

\makeatletter 
\def\subsubsection{\@startsection{subsubsection}{3}{10pt}{-1.25ex plus -1ex minus -.1ex}{0ex plus 0ex}{\normalsize\bf}} 
\def\paragraph{\@startsection{paragraph}{4}{10pt}{-1.25ex plus -1ex minus -.1ex}{0ex plus 0ex}{\normalsize\textit}} 
\renewcommand\@biblabel[1]{#1}            
\renewcommand\@makefntext[1]% 
{\noindent\makebox[0pt][r]{\@thefnmark\,}#1}
\makeatother 
\renewcommand{\figurename}{\small{Fig.}~}
\sectionfont{\large}
\subsectionfont{\normalsize} 

\fancyfoot{}
%arXiv \fancyfoot[LO,RE]{\vspace{-7pt}\includegraphics[height=9pt]{headers/LF}}
%arXiv \fancyfoot[CO]{\vspace{-7.2pt}\hspace{12.2cm}\includegraphics{headers/RF}}
%arXiv \fancyfoot[CE]{\vspace{-7.5pt}\hspace{-13.5cm}\includegraphics{headers/RF}}
%arXiv \fancyfoot[RO]{\footnotesize{\sffamily{1--\pageref{LastPage} ~\textbar  \hspace{2pt}\thepage}}}
%arXiv \fancyfoot[LE]{\footnotesize{\sffamily{\thepage~\textbar\hspace{3.45cm} 1--\pageref{LastPage}}}}
\fancyhead{}
\renewcommand{\headrulewidth}{1pt} 
\renewcommand{\footrulewidth}{1pt}
\setlength{\arrayrulewidth}{1pt}
\setlength{\columnsep}{6.5mm}
\setlength\bibsep{1pt}

\twocolumn[
  \begin{@twocolumnfalse}
\noindent\LARGE{\textbf{($4,4^\prime$)-Bipyridine in vacuo and in solvents: 
A quantum chemical study of a prototypical floppy molecule 
from a molecular transport perspective $^{\dag\P}$}}
\vspace{0.6cm}

\noindent\large{\textbf{Ioan B\^aldea,$^{\ast}$\textit{$^{a\ddag}$}
Horst K\"oppel,\textit{$^{a}$} and
Wolfgang Wenzel\textit{$^{b}$}}}\vspace{0.5cm}
%Please note that \ast indicates the corresponding author(s) but no footnote text is required. 

% \noindent\textit{\small{\textbf{Received Xth XXXXXXXXXX 20XX, Accepted Xth XXXXXXXXX 20XX\newline
% First published on the web Xth XXXXXXXXXX 200X}}}

\noindent \textbf{\small{DOI: 10.1039/C2CP43627B}}
\vspace{0.6cm}
%Please do not change this text.

\noindent \normalsize{
We report results of quantum chemical calculations for the neutral and anionic species of ($4,4^\prime$)-bipyridine (44BPY), a prototypical molecule with a floppy degree of freedom, placed in vacuo and in solvents. In addition to equilibrium geometries and vibrational frequencies and spectra, we present adiabatic energy curves for the vibrational modes with significant intramolecular reorganization upon charge transfer. Special attention is paid to the floppy strongly anharmonic degree of freedom of 44BPY, which is related to the most salient structural feature, namely the twist angle $\theta$ between the two pyridine rings. The relevance of the present results for molecular transport will be emphasized. We show that the solvent acts 
as a selective gate electrode and propose a scissor operator to account for solvent effects on molecular transport. Our result on the conductance $G$ vs. $\cos^2\theta$ is consistent with a significant transmission in  perpendicular conformation indicated by previous microscopic analysis.
}
\vspace{0.5cm}
 \end{@twocolumnfalse}
  ]

%Footnotes
\footnotetext{\dag~Electronic Supplementary Information (ESI) available: Additional details and tables. See DOI: 10.1039/C2CP43627B}
\footnotetext{\P~Published: Phys.\ Chem.\ Chem.\ Phys.\ 2013, {\bf 15}, 1918--1928}

%Please use \dag to cite the ESI in the main text of the article.
%If you article does not have ESI please remove the the \dag symbol from the title and the above footnotetext.

\footnotetext{\textit{$^{a}$~Theoretische Chemie, Universit\"at Heidelberg, Im Neuenheimer Feld 229, D-69120 Heidelberg, Germany.}}
\footnotetext{\ddag~E-mail: ioan@pci.uni-heidelberg.de. 
Also at National Institute for Lasers, Plasmas, and Radiation Physics, ISS, 
Bucharest, Romania}
\footnotetext{\textit{$^{b}$~Institut f\"ur Nanotechnologie, 
Karlsruher Institut f\"ur Technologie (KIT), 
D-76131 Karlsruhe, Germany}}

%%%%%%%%%%%%%%%%%%%%%%%%%%%%%%%%%%%%%%%%%%%%%%%%%%%%%%%%%%%%%%%%%
\section{Introduction}
\label{sec-intro}
%%%%%%%%%%%%%%%%%%%%%%%%%%%%%%%%%%%%%%%%%%%%%%%%%%%%%%%%%%%%%%%%%%%%%%%%%%%%%%%%%%%%%%%%%%%%%%% 
The present work represents a quantum chemical study intended to pave the way toward 
microscopic nanotransport studies in single-molecule devices 
in vacuo and in solvents based on (4, 4$^\prime$)-bipyridine (44BPY),
a prototypical floppy molecule,
wherein a floppy vibrational mode turns out to yield an important
intramolecular reorganization upon charge transfer. This is an effect,
which existing transport studies did not consider.

In view of its special structure, 44BPY is a representative model molecule 
that attracted considerable experimental and theoretical interest. 
Compounds based on 44BPY, commonly known as viologens, captured chemists' 
attention for many decades, especially due to their electrochemically reversible 
behavior.\cite{Bird:81,Monk:01} They have been investigated 
from various perspectives, like
herbicidal activity,\cite{Summers:81} electrochemical display devices, 
storage energy, catalytic oxidation, 
\cite{Summers:81,Graetzel:81,Knopps:94,Vitale:99}
photoactive molecular, and supramolecular machines.\cite{Saha:07}
The preparation of 
complexes embedding neutral, radical anionic and dianionic forms of 
(4, 4$^\prime$)-bipyridyl ligands is of interest for designing switchable materials 
wherein the redox chemistry can be finely matched to facilitate electron transfer.
% \cite{Goicoechea:11}.

More recently, the 44BPY molecule has been used in 
molecular electronics. \cite{Cunha:96,Tao:03,Wandlowski:08}
Due to its special structure, 
% (4, 4$^\prime$)-bipyridine (44BPY) is a
% well suited molecule applications involving electron (photo)transfer mechanisms.
with two active nitrogen atoms in para positions (cf.~Fig.~\ref{fig:44bpy}), 
44BPY is particularly suitable for simultaneous binding to metallic electrodes. 
This fact 
has been exploited in the first work that succeeded in the repeated formation of 
numerous molecular junctions.\cite{Tao:03} 
44BPY molecules have been incorporated in 
redox active tunneling junctions to demonstrate the 
electrolyte gating.\cite{Wandlowski:08}

It is from the perspective of molecular 
electron transport that the present quantum chemical study has been conducted.
Unlike in most molecular electronic devices fabricated so far, in the single-molecule 
junctions based on 44BPY there exists clear experimental evidence that the 
electric current is due to negatively charged carriers (n-type conduction):
the measured Seebeck coefficient is negative.\cite{Venkataraman:12}
This indicates that the electron tunneling through molecular junctions wherein a 44BPY 
molecule is
contacted to gold electrodes is predominantly due to the lowest unoccupied molecular orbital (LUMO).
A positive Seebeck coefficient, with the associated p-type conduction, would have revealed 
the predominance of the highest occupied molecular orbital (HOMO). 
As a further support, one can invoke the results on
electrochemical gate-controlled electron transport through redox-active 
viologen molecular junctions,\cite{Wandlowski:08} which also indicate a LUMO-mediated conduction. 
This kind of experiments also enables one to distinguish between
electron (LUMO-mediated) n-type conduction and hole (HOMO-mediated) p-type conduction: 
the gate potential (overpotential in electrochemists' nomenclature) shifts 
the energies of the occupied and unoccupied molecular orbital in opposite directions relative to 
the electrodes' Fermi level, enhancing or diminishing the electric current.

By implication, this means that, when an electron subjected to a source-drain voltage 
travels across a 44BPY-based molecular junction, a transient radical anion (44BPY$^{\bullet -}$)
is formed. Therefore, the determination of the properties of the neutral 44BPY$^0$ molecule as 
well as of the bipyridyl radical 44BPY$^{\bullet -}$ is a prerequisite for understanding 
the electron transport in these molecular junctions. 
This is the aim of the present paper.
% , which represents
% a complementary study to a recent investigation done by one of us \cite{Baldea:2012i}.

% \section{Methodology}
The quantum chemical calculations reported below have been performed by using the Gaussian 09
suite of programs \cite{g09} at DFT/B3LYP level. 
Our study substantially goes beyond earlier works  
wherein both the parent molecule 44BPY$^{0}$  and the radical anion 44BPY$^{\bullet -}$ 
have been investigated theoretically, the latter  
especially in relation with 
photoreduction.\cite{Barone:83,Barone:85,Kihara:86,Poizat:91b,Kassab:96,Kassab:98,Perez:05b,Zhuang:07}
The aforementioned theoretical works used smaller basis sets and only considered 
a few ground state properties of
44BPY$^{0}$ and 44BPY$^{\bullet -}$ in vacuo.
In addition to the case of vacuum, we also present results for 
44BPY in solvents of experimental interest.
% To perform such calculations, the solvent has been described 
The solvent has been described
within the polarized continuum model using the integral equation formalism
(keyword SCRF=IEFPCM in Gaussian 09).
\ib{For the anionic (and cationic, cf.~Sec.~\ref{sec-solvent})
species, we carried out unrestricted calculations (Gaussian keyword UB3LYP).
Because calculations for open shell systems can sometimes be problematic 
(wrong geometries, spurious symmetry breaking,
or spin contamination, see, e.~g., Ref.~\citenum{AyanDatta:11}), 
to avoid artefacts we always checked our results carefully,
and convinced ourselves that, for instance, optimizations properly converged
(all frequencies are real),
symmetries (point group D$_{2}$ for 44BPY$^{0}$ and 44BPY$^{\bullet +}$,
and D$_{2h}$ for 44BPY$^{\bullet -}$)
and spin are correct. For concreteness, we mention that,
with the default settings of Gaussian 09, typical spin values found
are $\langle S^2 \rangle = 0.7592$ before annihilation of the first contaminant 
and $\langle S^2 \rangle = 0.7501$ after annihilation. To be complete, let us still note that no
problem arose even 
in the ``small'' calculation for the anion done at the very modest 
ROHF/6-31g(+*) level of theory (cf.~end of Sec.~\ref{sec-frequencies}); that calculation 
has been solely done to allow comparison with and check results of 
earlier work.}

The remaining part of this paper is organized as follows.
In Section \ref{sec-geometry}, 
results for the equilibrium geometries of the neutral and anionic species
obtained by using basis sets larger than those employed in previous studies
will be reported.
Further, vibrational effects in the radical anion placed in vacuo and in experimental conditions
(acetonitrile as solvent \cite{Kassab:96}) will be discussed in 
Section \ref{sec-frequencies}.
Section \ref{sec-adiabatic-curves} 
is devoted to an analysis of
the relevant adiabatic energy curves for 44BPY$^{0}$ and 44BPY$^{\bullet -}$; this analysis
is augmented with the discussion in the Electronic Supplementary Information. 
Results for the 44BPY molecule immersed in
solvents will be presented in 
Section \ref{sec-solvent}.
Because the present quantum chemical investigation is intended to pave the way toward
transport studies in 44BPY-based molecular devices,
we present in 
Section \ref{sec-G} 
preliminary
results revealing the impact of the molecular conformation on the electrical conductance.
Discussions and conclusions will be presented in
Section \ref{sec-conclusion}.
% section 7.
%%%%%%%%%%%%%%%%%%%%%%%%%%%%%%%%%%%%%%%%%%%%%%%%%%%%%%%%%%%%%%%%%%%%%%%%%%%%%%%%%%%%%%%%%%%%%%%%%%%%%%%
% \section{Results}
%%%%%%%%%%%%%%%%%%%%%%%%%%%%%%%%%%%%%%%%%%%%%%%%%%%%%%%%%%%%%%%%%%%%%%%%%%%%%%%%%%%%%%%%%%
\section{Equilibrium geomerties of 44BPY$^0$ and 44BPY$^{\bullet -}$}
\label{sec-geometry}
%%%%%%%%%%%%%%%%%%%%%%%%%%%%%%%%%%%%%%%%%%%%%%%%%%%%%%%%%%%%%%%%%%%%%%%%%%%%%%%%%%%%%%%%%%%
In this study, quantum chemical calculations for geometry optimizations without imposing 
any constraints have been performed 
both for the neutral and for the anionic species. 
In such calculations, an important issue is to employ a sufficiently large basis set.

\begin{figure}[h!]
$ $\\[6ex]
% 
% \centerline{\hspace*{-0ex}\includegraphics[width=0.4\textwidth,angle=0]{44bpyCP.eps}}
\centerline{\hspace*{-0ex}\includegraphics[width=0.4\textwidth,angle=0]{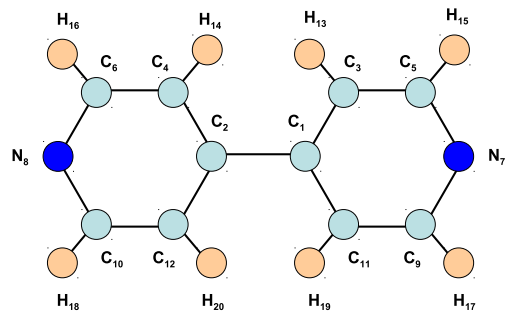}}
$ $\\[6ex]
\caption{The ($4, 4^\prime$) molecule (44BPY) and the atom numbering used in 
the present work.}
\label{fig:44bpy}
\end{figure}

We have extensively investigated the impact of the basis set size.
Table S1 (tables with label S are from the Electronic Supplementary Information) % Table \ref{table-neutral} 
collects our results on the equilibrium geometry of the neutral 
44BPY molecule obtained via DFT/B3LYP and various basis sets. 
The atom numbering is given in Fig.~\ref{fig:44bpy}.
%%%%%%%%%%%%%%%%%%%%%%%%%%%%%%%%%%%%%%%%%%%%%%%%%%%%%%%%%%%%%%%%%%%%%%%%%%%%%%%%%%%%%%%%%%
%%%%%%%%%%%%%%%%%%%%%%%%%%%%%%%%%%%%%%%%%%%%%%%%%%%%%%%%%%%%%%%%%%%%%%%%%%%%%%%%%%%%%%%
The optimum geometry of the neutral species has been previously investigated in a series
of studies.\cite{Barone:83,Barone:85,Kassab:96,Kassab:98,Kassab:05,Zhuang:07}
These studies used lower theoretical levels (semi-empirical, HF) and/or smaller basis sets 
(sto-3g, 3-21g(+*), 6-31g(+*) \cite{notationPlusAst}).
% where the notation ``(+*)'' means that 
% diffuse and polarization functions have been added only to the nitrogen atoms).
As compared to those results, our results represent an
improved agreement with experiment.\cite{Almenningen:58,Mata:77}
For the cc-pVTZ basis set, the largest set used earlier for 
44BPY$^0$,\cite{Perez:05b} our study basically confirms the results reported in 
Ref.~\citenum{Perez:05b}. We note that there exist small
differences between the present geometrical parameter
values computed with Gaussian 09 and those (also included in Table S1) % Table \ref{table-neutral}) 
computed with GAMESS in Ref.~\onlinecite{Perez:05b}
at the same DFT/B3LYP/cc-pVTZ level of theory. However, they are small; 
bond lengths differ by at most 
0.001\,{\AA}. By and large, Table S1 % Table \ref{table-neutral} 
confirms the previous claim 
\cite{Kassab:98} that the geometrical parameters of 44BPY$^0$ are little sensitive 
to the basis set size.
An important fact to be emphasized in connection to Table S1 % Table \ref{table-neutral}
is that, while substantially increasing the computational cost, the usage of larger 
basis sets does not necessarily improve the agreement between theoretical and experimental
bond metric data.
For instance, although 
the inter-ring C--C bond length is somewhat reduced by increasing the basis set, 
it remains slightly 
larger than the experimental one.\cite{Mata:77} The theoretical value becomes closer 
to a normal C(sp$^2$)--C(sp$^2$) single bond (1.479\,{\AA}).\cite{Dewar:58}

Most sensitive to the basis set size is the twisting angle 
$\theta = \widehat{ C_3 C_1 C_2 C_4} $ 
between the two pyridyl rings, which has been long recognized to be the salient structural
feature of this molecule,\cite{Barone:83,Barone:85,Kassab:96,Kassab:98} 
whose variations are related to very small energy variations. 
%%%%%%%%%%%%%%%%%%%%%%%%%%%%%%%%%%%%%%%%%%%%%%%%%%%%%%%%%%%%%%%%%%%%%%%%%%%%%%%%%%%%%%%%%

Our bond metric data for the radical anion 44BPY$^{\bullet -}$ are shown in 
Table S2. % Table~\ref{table-anion}. 
One should note that the previous theoretical studies on the anion 
have been done either semi-empirically \cite{Kihara:86}
or by using much smaller basis sets (HF/3-21g(+*) \cite{Kassab:96}
and DFT/6-31g(+*).\cite{Kassab:98,notationPlusAst}
% \cite{Kihara:86,Kassab:96,Kassab:98,Lapouge:02}.
In the absence of direct experimental results for the anion structure, we can only 
present the comparison with existing theoretical results obtained for the largest basis,
namely 6-31 g(+*) of Ref.~\onlinecite{Kassab:98}. 
Somewhat surprisingly, 
the geometrical parameters are little sensitive 
to the basis set size even in the case of the radical anion 44BPY$^{\bullet -}$. 
Noteworthy, 
the computed geometrical parameters 
are rather insensitive to the presence of the 
diffuse basis functions. This behavior contrasts with that exhibited by 
other properties of the anion, which will be
examined later in Sections \ref{sec-adiabatic-curves} and \ref{sec-G}.
%%%%%%%%%%%%%%%%%%%%%%%%%%%%%%%%%%%%%%%%%%%%%%%%%%%%%%%%%%%%%%%%%%%%%%%%%%%%%%%%%%%%%%%%%

Because the structural differences between the neutral molecule 44BPY$^0$ 
and its reduced form 44BPY$^{\bullet -}$ have been amply discussed in the literature (see, e.~g., 
Refs.~\onlinecite{Kassab:96,Kassab:98} and \onlinecite{Baldea:2012i}), 
we only briefly mention the main aspects.
The twisting angle $\theta$
% $\theta = \widehat{ C_3 C_1 C_2 C_4} $ 
between the two pyridyl rings results from the interplay between the $\pi$-electronic 
interaction of the pyridyl fragments and the steric repulsion
of the ortho hydrogen atoms (H$_{13}$ and H$_{14}$, and H$_{19}$ and H$_{20}$ in Fig.~\ref{fig:44bpy}).
% with respect to the C--C inter-ring bond. 
The former favors
a planar conformation ($\theta = 0$) wherein the $\pi$-electrons are delocalized 
between the pyridyl rings, while the latter prefers a nonplanar 
conformation ($\theta \neq 0$) wherein the steric hindrance is reduced.
The neutral molecule can be characterized as consisting of two aromatic pyridyl 
rings linked by a (nearly) single C--C bond in a twisted conformation (point group D$_2$).
The balance between the two aforementioned effects in the neutral molecule yields an 
equilibrium value $\theta_{eq} = 37.2^\circ$, as measured in electron diffraction 
experiments,\cite{Almenningen:58} which can be well reproduced by quantum chemical calculations 
(cf.~Table S1). % Table \ref{table-neutral}). 

As suggested for the first time on the basis of a semi-empirical MO 
approach,\cite{Kihara:86} and later within HF/3-21g(+*) \cite{Kassab:96} 
and DFT/6-31g(+*) \cite{Kassab:98} calculations, two
important structural changes occur on going from 
the parent molecule 44BPY$^0$ to the radical anion 44BPY$^{\bullet -}$. 
First, the $\pi$-electronic interaction is enhanced by the extra electron 
delocalized over the reduced molecule, 
which becomes planar ($\theta_{eq} = 0$, point group D$_{2h}$).
Second, the anion is characterized by a (planar) quinoidal structure.
Basically, the quinoidal distortion is localized in the inter-ring region. 
On going from the neutral to the reduced species, mostly affected 
are the inter-ring C$_1$--C$_2$ bond length, which shortens by $\sim 0.052$\,{\AA},
becoming thereby substantially stronger and acquiring partial double bond character
(bond order 1.247, see below),
and the neighboring C$_1$--C$_3$ bond length, which increases by $\sim 0.034$\,{\AA}.
Less affected are the C$_5$--N$_7$ bond, which lengthens by $\sim 0.018$\,{\AA},
and the C$_3$--C$_5$ ring bond, which shortens by $\sim 0.012$\,{\AA}. The calculated 
bond orders are only slightly dependent on the method. With ROHF/6-31g(+*) we found 1.288 and 1.532 
for C$_1$--C$_2$- and C$_3$--C$_5$-(Wiberg) bond order indices, 
confirming thereby the previously reported
values 1.29 and 1.53, respectively;\cite{Lapouge:02} B3LYP/aug-cc-pVTZ yields the values
1.247 and 1.560, respectively.\cite{Annika}
%%%%%%%%%%%%%%%%%%%%%%%%%%%%%%%%%%%%%%%%%%%%%%%%%%%%%%%%%%%%%%%%%%%%%%%%%%%%%%%%%%%%%%%%%%%%%%% 
\section{Vibrational modes of the radical anion 44BPY$^{\bullet -}$}
\label{sec-frequencies}
%%%%%%%%%%%%%%%%%%%%%%%%%%%%%%%%%%%%%%%%%%%%%%%%%%%%%%%%%%%%%%%%%%%%%%%%%%%%%%%%%%%%%%%%%
For the neutral 44BPY molecule, theoretical results obtained by using DFT/B3LYP and 
good basis sets 
(e.~g., cc-pVTZ \cite{Perez:05b}) 
% (cc-pVTZ \cite{Perez:05b} and 6-31++g(d, p) \cite{Zhuang:07}) 
have already been compared 
with the experimental data available for the Raman 
\cite{Kassab:96} and infra-red (IR) active \cite{Kassab:96,Zhuang:07}
vibrational modes. By using basis sets larger (cf.~Table S1) % Table \ref{table-neutral})
than that employed in Ref.~\onlinecite{Perez:05b}, we did not
find a significant improvement against experiments: % \cite{Baldea:unpublished}: 
the empirical factors, which are commonly introduced
\cite{Scott:96,Sinha:04} to scale the calculated vibrational frequencies,
remain close to the value $\sim 0.98$ found in Ref.~\onlinecite{Perez:05b}.

\begin{figure}[h!]
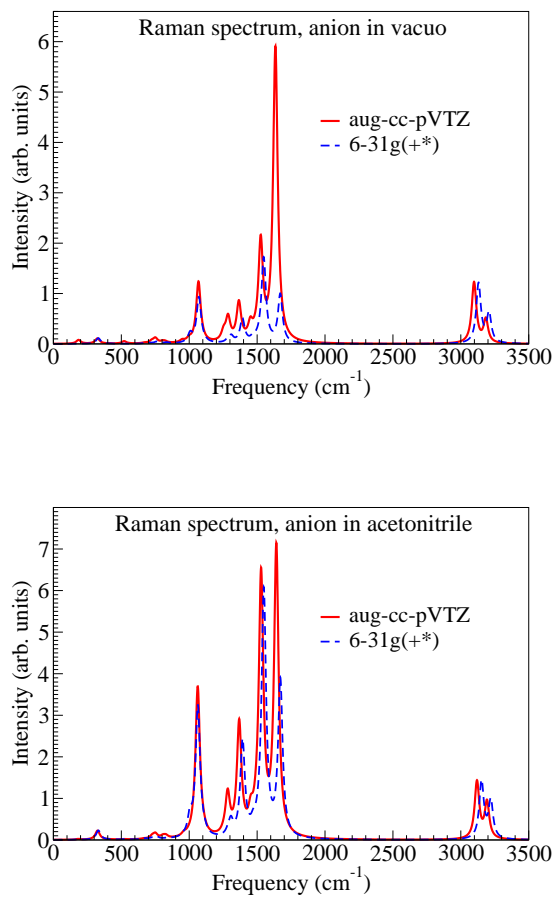

$ $\\[6ex]
% 
% \centerline{\hspace*{-0ex}\includegraphics[width=0.4\textwidth,angle=0]{RamanSpectrum44bpyAnion_lorentzian_hw_20cm_1.eps}}
\centerline{\hspace*{-0ex}\includegraphics[width=0.4\textwidth,angle=0]{Fig2a.eps}}
$ $\\[6ex]
% \centerline{\hspace*{-0ex}\includegraphics[width=0.4\textwidth,angle=0]{RamanSpectrum44bpyAnionAcetonitrile_lorentzian_hw_20cm_1.eps}}
\centerline{\hspace*{-0ex}\includegraphics[width=0.4\textwidth,angle=0]{Fig2b.eps}}
$ $\\[6ex]
\caption{Theoretical Raman spectra 
calculated with the basis sets 6-31g(+*) \cite{notationPlusAst} and aug-cc-pVTZ 
for the radical anion 44BPY$^{-1}$ placed in vacuo and acetonitrile. 
They illustrate that both
the basis set size and the solvent significantly affect the Raman spectral intensities. 
The computed 
spectral lines have been convoluted with Lorentzian functions of half-width 20\,cm$^{-1}$.
See the main text for details.}
\label{fig:raman}
\end{figure}

\begin{figure}[h!]
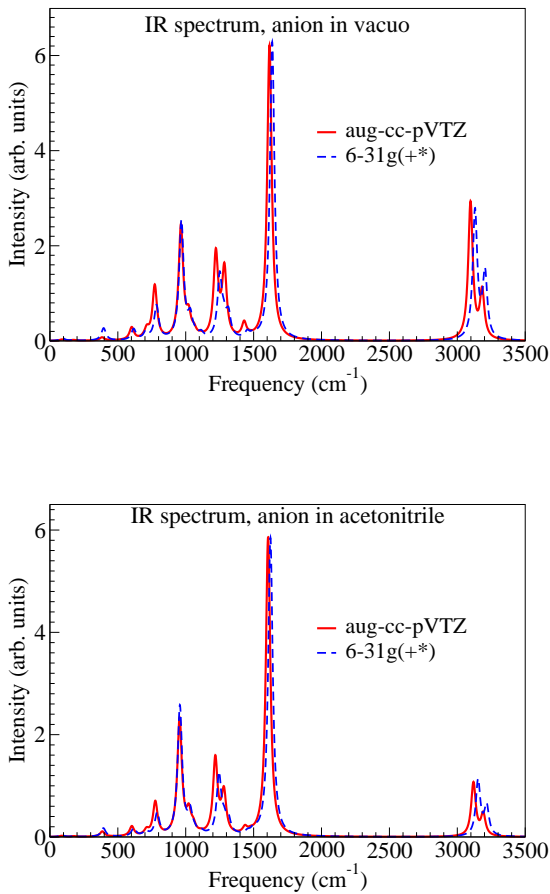

$ $\\[6ex]
% 
% \centerline{\hspace*{-0ex}\includegraphics[width=0.4\textwidth,angle=0]{IRSpectrum44bpyAnion_lorentzian_hw_20cm_1.eps}}
\centerline{\hspace*{-0ex}\includegraphics[width=0.4\textwidth,angle=0]{Fig3a.eps}}
$ $\\[6ex]
% \centerline{\hspace*{-0ex}\includegraphics[width=0.4\textwidth,angle=0]{IRSpectrum44bpyAnionAcetonitrile_lorentzian_hw_20cm_1.eps}}
\centerline{\hspace*{-0ex}\includegraphics[width=0.4\textwidth,angle=0]{Fig3b.eps}}
$ $\\[6ex]
\caption{Theoretical IR spectra 
calculated with the basis sets 6-31g(+*) \cite{notationPlusAst} and aug-cc-pVTZ 
for the radical anion 44BPY$^{-1}$ placed in vacuo and acetonitrile.
They illustrate that
neither the basis set size nor the solvent drastically affect the spectral intensities.
The computed 
spectral lines have been convoluted with Lorentzian functions of half-width 20\,cm$^{-1}$ 
See the main text for details.}
\label{fig:ir}
\end{figure}

Therefore, we confine ourselves to discuss vibrational effects for the 
radical anion,
wherein earlier results \cite{Kassab:96,Kassab:98} have been obtained at 
modest theoretical levels (HF/3-21g(+*) \cite{Kassab:96} and DFT/B3LYP/6-31G(+*).\cite{Kassab:98})
The results for the frequencies of the Raman active modes, 
which can be compared with available experimental data,\cite{Kassab:96} 
are collected in Table S3. % \ref{table-freq-anion}. 
Although Raman measurements have also been carried out for 
the partially or totally deuterated radical anion \cite{Kassab:96}), 
we confine ourselves to the undeuterated case, of more direct relevance for 
electron transport. The Raman measurements \cite{Kassab:96} comprise the
spectral range $300$\,cm$^{-1} < \omega < 2000$\,cm$^{-1}$. 
It does not include the spectral range $\omega > 3000$\,cm$^{-1}$ wherein C--H stretching modes 
are active (see Figs.~\ref{fig:raman} and \ref{fig:ir}). 
Out of the eight Raman active modes (A$_{g}$ symmetry) shown in Table S3, % \ref{table-freq-anion},
only seven have been detected experimentally.\cite{Kassab:96} The lowest ring deformation mode
($6a$, 
% cf.~Table \ref{table-freq-anion}) 
cf.~Table S3) 
has not been observed. 
This can be understood: the spectra calculated by us 
(Fig.~\ref{fig:raman}) show indeed a very weak spectral intensity for this mode.  

As compared with the 
approach HF/3-21g(+*) of Ref.~\onlinecite{Kassab:96}, where the empirical scaling factor 
amounts 0.892, the approach of Ref.~\onlinecite{Kassab:98} already represented a substantial 
improvement: the empirical scaling factor needed there was substantially larger ($f_{sc}=0.968$). 
The agreement is further improved by using larger basis sets, as expressed
by the values $f_{sc} \simeq 0.98$ of 
% Table~\ref{table-freq-anion}. 
Table~S3. 
So, the agreement is 
basically as good as that for the neutral molecule.
To conclude, the DFT approach is able to accurately reproduce experimental vibrational frequencies:
the deviations amount to $\sim 2$\%; for the modes shown in Table S3 % \ref{table-freq-anion} 
this represents an absolute deviation of $\sim 10 - 30$\,cm$^{-1}$.
One should still note that, similar to the case of the geometrical parameters, 
further improvement cannot be achieved by employing larger basis
sets (for aug-cc-pVQZ, we found $f_{sc}=0.980$), a fact which reflects the intrinsic limitation of the 
current theory.

In contrast to the fact, already noted, of relatively insensitive 
vibrational frequencies,
the Raman intensities are much more sensitive to the basis set. 
This is exemplified by the Raman
spectrum presented in Fig.~\ref{fig:raman}.\cite{not-for-neutral}

Because the Raman measurements refer to the anion in solvent (acetonitrile),\cite{Kassab:96}
we have also performed calculations for this situation. 
% As also the case in Sec.~\ref{sec-solvent}, the solvent has been described ``hopa''
% within the polarized continuum model using the integral equation formalism
% (keyword SCRF=IEFPCM in Gaussian 09). 
As visible in 
% Table~\ref{table-freq-anion} 
Table~S3
and
Fig.~\ref{fig:raman}, while weakly affecting the vibrational frequencies, the solvent 
has a significant impact on the spectral Raman intensities. 
%%%%%%%%%%%%%%%%%%%%%%%%%%%%%%%%%%%%%%%%%%%%%%%%%%%%%%%%%%%%%%%%%%%%%%%%%%%%%%%%%%%%%%%%%%
%%%%%%%%%%%%%%%%%%%%%%%%%%%%%%%%%%%%%%%%%%%%%%%%%%%%%%%%%%%%%%%%%%%%%%%%%%%%%%%%%%%%%%%

Although we are not aware of experimental 
studies on vibrational spectra of the infra-red (IR) active modes in the radical anion 44BPY$^{\bullet -}$, 
for completeness 
we have also calculated the IR spectrum. These results are presented in
Fig.~\ref{fig:ir}. As compared to the Raman spectra, the IR spectra depicted in
Fig.~\ref{fig:ir}
turn out to be much less affected both by the basis set size and by the presence of the solvent.
%%%%%%%%%%%%%%%%%%%%%%%%%%%%%%%%%%%%%%%%%%%%%%%%%%%%%%%%%%%%%%%%%%%%%%%%%%%%%%%%%%%%%%%%%%
%%%%%%%%%%%%%%%%%%%%%%%%%%%%%%%%%%%%%%%%%%%%%%%%%%%%%%%%%%%%%%%%%%%%%%%%%%%%%%%%%%%%%%%
%%%%%%%%%%%%%%%%%%%%%%%%%%%%%%%%%%%%%%%%%%%%%%%%%%%%%%%%%%%%%%%%%%%%%%%%%%%%%%%%%%%%%%%
\section{Adiabatic ground state energy curves of 44BPY$^0$  and 44BPY$^{\bullet -}$}
\label{sec-adiabatic-curves}
%%%%%%%%%%%%%%%%%%%%%%%%%%%%%%%%%%%%%%%%%%%%%%%%%%%%%%%%%%%%%%%%%%%%%%%%%%%%%%%%%%%%%%%
The coherent electron transport through single-molecule 
junctions based on 44BPY was investigated theoretically 
in several works.\cite{Hou:05,Thygesen:05c,Bagrets:08,Venkataraman:12}
In spite of significant differences between them, there is a common finding in those studies, namely 
that the coherent electron transport is determined by the LUMO: 
close to the metallic Fermi energy, the calculated 
transmission exhibits a single pronounced Lorentzian peak 
and the Seebeck coefficient is negative.
The LUMO energy can be obtained 
from electronic structure $\Delta$-DFT 
calculations: it represents the difference between the anion and 
neutral ground state energies $E_{LUMO} = E_{A}  - E_{N}$ 
at the neutral equilibrium geometry $\mathbf{Q}_N$.\cite{lumo} 

Refs.~\onlinecite{Hou:05,Thygesen:05c,Bagrets:08,Venkataraman:12}
assumed a molecular geometry frozen at 
$\mathbf{Q} = \mathbf{Q}_N$($\equiv \mathbf{0}$).
Our recent study \cite{Baldea:2012i} drew attention to the fact that intramolecular 
reorganization is important and should be considered, especially because the 44BPY molecule 
possesses a floppy degree of freedom. 
In particular, this implies that the LUMO energy is not fixed at the value
$E_{LUMO} = E_{A}(\mathbf{Q}_N)  - E_{N}(\mathbf{Q}_N)$.
Instead, one should consider a distribution of values
$E_{LUMO}(\mathbf{Q}) = E_{A}(\mathbf{Q}) - E_{N}(\mathbf{Q})$,
and perform a $\mathbf{Q}$-ensemble averaging of the current computed 
at various molecular geometries $\mathbf{Q}$. The weight function needed for this 
ensemble averaging requires the knowledge of the adiabatic Gibbs free energy 
surface $\mathcal{G}(\mathbf{Q})$. The calculation of $\mathcal{G}(\mathbf{Q})$ is nontrivial 
(cf., e.~g., Refs.~\onlinecite{Zhang:05,Baldea:2012i} and citations therein).
Necessary (but not sufficient)  
quantities that should be determined from electronic structure calculations 
are the adiabatic potential 
energy surfaces $E_{N,A}(\mathbf{Q})$ of the neutral and anion electronic 
ground states at arbitrary geometries $\mathbf{Q}$, as has been done in the
present work.

Applied to the case considered here, intramolecular reorganization 
manifests itself in the different locations $\mathbf{Q}_A \neq \mathbf{Q}_N$ 
of the minima of the adiabatic energy surfaces $E_{A}(\mathbf{Q})$ and $E_{N}(\mathbf{Q})$,
respectively. 
The reorganization energies $\lambda_{N,A}$ of the neutral and anionic species 
represent important quantities for electron transport both in the nonadiabatic 
(Mulliken-Hush-Marcus-type) \cite{Marcus:85,Marcus:93,Ratner:00} and adiabatic 
\cite{Schmickler:93,Kuznetsov:02c,Zhang:05,Medvedev:07,Baldea:2012i}
limits. They are defined as 
%%%%%%%%%%%%%%%%%%%%%%%%%%%%%%%%%%%%%%%%%%%%%%%%%%%%%%%%%%%%%%%%%%%%%%%%%%%%%%%%%%%%%%%%
\begin{eqnarray}
\lambda_{N} & = & E_{N}(\mathbf{Q}_A) - E_{N}(\mathbf{Q}_N) , \nonumber \\
\lambda_{A} & = & E_{A}(\mathbf{Q}_N) - E_{A}(\mathbf{Q}_A) .
\label{eq-lambda}
\end{eqnarray}
%%%%%%%%%%%%%%%%%%%%%%%%%%%%%%%%%%%%%%%%%%%%%%%%%%%%%%%%%%%%%%%%%%%%%%%%%%%%%%%%%%%%%%%%

\begin{figure}[h!]
$ $\\[0.6ex]
% 
% \centerline{\hspace*{-0ex}\includegraphics[width=0.38\textwidth,angle=0]{figs/fig_Q8a.eps}}
\centerline{\hspace*{-0ex}\includegraphics[width=0.38\textwidth,angle=0]{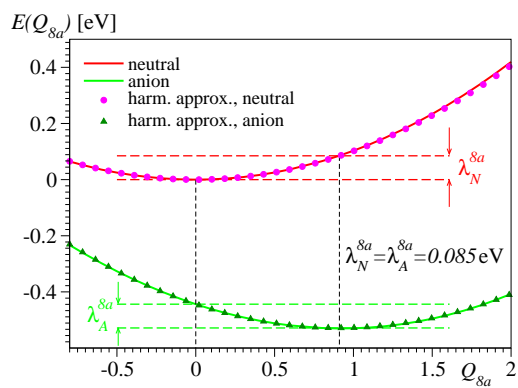}}
$ $\\[0.6ex]
\caption{Adiabatic energy curves along the coordinate $Q_{8a}$ of mode 8a 
of the neutral molecule 
($\omega_{8a}^{N} = 1640$\,cm$^{-1}$) computed with the basis set aug-cc-pVDZ 
for neutral 
44BPY and radical anion bipyridyl 44BPY$^{\bullet -}$. Notice the almost 
perfect harmonic behavior and the fact that the 
neutral and anion minima are located on the same side with respect to the intersection 
point of the curves, situated at left beyond the figure's frame (inverted regime).}
\label{fig:e-Q-8a}
\end{figure}

Our calculations confirm the intuitive expectation that significant contributions to the 
reorganization energy arise from the totally symmetric in-plane normal modes. 
The Raman active vibrational modes included in 
Table S3 % \ref{table-freq-anion} 
belong to these modes.
Out of them, the largest contribution to the reorganization energy 
comes from the mode 8a. 
Adiabatic energy curves $E_{N,A}(Q_{8a})$ along the normal coordinate $Q_{8a}$ 
(all the other $\mathbf{Q}$'s being set to zero)
are presented in Fig.~\ref{fig:e-Q-8a} and indicate a virtually perfect harmonic behavior.

Considered alone, each of the Raman active modes other than $8a$ 
listed in Table S3 % \ref{table-freq-anion} 
has a small contribution to reorganization and also exhibits a harmonic behavior. 
Their contribution can be accounted for by a renormalization factor 
of mode $8a$
$f_{h}$ ($\omega_{8a} \to \tilde{\omega}_h  = f_{h}\,\omega_{8a}$), 
as discussed in the Electronic Supplementary Information.\cite{Schmickler:96b}
In accord with a general property of the harmonic modes,\cite{Baldea:2012i}
the corresponding partial reorganization energies for the neutral and anionic species 
should be equal. For the mode $8a$ alone, we found
$\lambda_{N}^{8a} = \lambda_{A}^{8a} = 0.0845$\,eV. 
%%%%%%%%%%%%%%%%%%%%%%%%%%%%%%%%%%%%%%%%%%%%%%%%%%%%%%%%%%%%%%%%%%%%%%%%%%%%%%%%%%%%%%%
%%%%%%%%%%%%%%%%%%%%%%%%%%%%%%%%%%%%%%%%%%%%%%%%%%%%%%%%%%%%%%%%%%%%%%%%%%%%%%%%%%%%%%

\begin{figure}[h!]
$ $\\[0.6ex]
% 
% \centerline{\hspace*{-0ex}\includegraphics[width=0.38\textwidth,angle=0]{figs/fig_Q0.eps}}
\centerline{\hspace*{-0ex}\includegraphics[width=0.38\textwidth,angle=0]{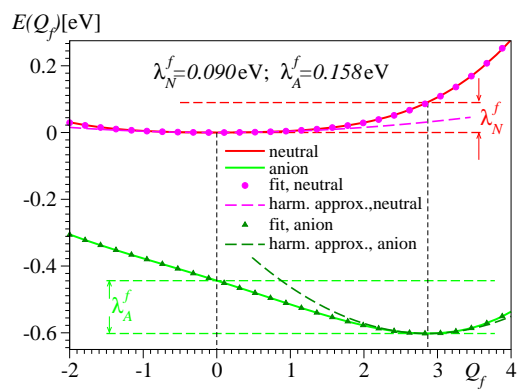}}
$ $\\[0.6ex]
\caption{Adiabatic energy curves along the coordinate $Q_{f}$ of the neutral molecule 
of the lowest frequency mode 
($\omega_f = 62$\,cm$^{-1}$) computed with the basis set aug-cc-pVDZ for neutral 
44BPY and radical anion bipyridyl 44BPY$^{\bullet -}$. Notice the strong departures from  
curves corresponding to the harmonic approximation.
See the main text for details.}
\label{fig:e-Q-f}
\end{figure}

However, the values of the total reorganization energies 
$\lambda_{N,A}$ computed from Eq.~\ref{eq-lambda} were found different,
$\lambda_N \simeq 0.23$\,eV  and $\lambda_A \simeq 0.35$\,eV, and this inequality traces back 
to the lowest frequency mode, which is strongly anharmonic.\cite{Baldea:2012i}
This mode is directly related to the inter-ring torsional motion and
represents the floppy (label $f$) degree of freedom of the 44BPY molecule.\cite{Baldea:2012i}
Its frequency $\omega_f \simeq 62$\,cm$^{-1}$ \cite{Baldea:2012i} lies well below the 
spectral vibrational range investigated experimentally 
($400$\,cm$^{-1} \alt \omega \alt 1800$\,cm$^{-1}$) for 44BPY$^{0}$ \cite{Zhuang:07} and
$300$\,cm$^{-1} \alt \omega \alt 2000$\,cm$^{-1}$) for 44BPY$^{\bullet -}$.\cite{Kassab:96}
However, in spite of this very low frequency, its contribution to 
the reorganization energy is important because of the large amplitude vibrations associated with it.
The anionic and neutral adiabatic energy curves along the coordinate $Q_f$ of the neutral molecule, 
which are depicted 
in Fig.~\ref{fig:e-Q-f}, exhibit strong anharmonicities. 
The fact that strong anharmonic effects characterize molecules possessing floppy degrees 
of freedom is well known (see, e.~g., Ref.~\citenum{Tucker:94}). 
However, except for the preliminary 
work of Ref.~\onlinecite{Baldea:2012i}, 
we are not aware of (transport-related) studies drawing attention to the fact that 
such modes can be strongly
coupled to molecular orbitals with essential contribution to the charge transfer.    
%%%%%%%%%%%%%%%%%%%%%%%%%%%%%%%%%%%%%%%%%%%%%%%%%%%%%%%%%%%%%%%%%%%%%%%%%%%%%%%%%%%%%%%
%%%%%%%%%%%%%%%%%%%%%%%%%%%%%%%%%%%%%%%%%%%%%%%%%%%%%%%%%%%%%%%%%%%%%%%%%%%%%%%%%%%%%% 

\begin{figure}[h!]
$ $\\[0.6ex]
% 
% \centerline{\hspace*{-0ex}\includegraphics[width=0.38\textwidth,angle=0]{figs/fig_Q0_basis_sets.eps}}
\centerline{\hspace*{-0ex}\includegraphics[width=0.38\textwidth,angle=0]{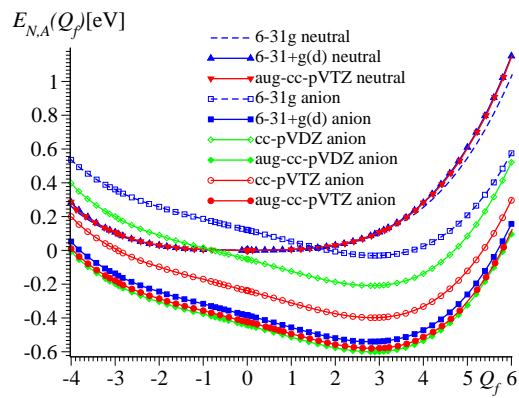}}
$ $\\[0.6ex]
\caption{The adiabatic energy curves $E_{N,A}(Q_f)$ of the neutral 44BPY$^0$ 
and anion species along the coordinate $Q_f$ of the floppy degree of freedom
computed for several basis sets. 
\ib{For the neutral species, the curves computed with the basis sets 6-31+g(d) and aug-cc-pVTZ are practically indistinguishable.}
Notice the fact that, unlike $E_{N}(Q_f)$,
rather than the basis set size, the inclusion of diffuse basis functions 
is essential 
to correctly compute the adiabatic energy $E_{A}(Q_f)$ of 44BPY$^{\bullet -}$. 
Results obtained with the modest 6-31+g(d) basis set are reasonably good,
while those with the basis set cc-pVTZ are not.
See the main text for details.}
\label{fig:e-Q-f-basis}
\end{figure}

In Fig.~\ref{fig:e-Q-f-basis}, we present adiabatic curves $E_{N,A}(Q_f)$
for the neutral and anionic species calculated with various basis sets.
The inspection of these curves reveals an important difference between the neutral and  
anionic species. 
The adiabatic curves of the neutral molecule are weakly affected by the basis set employed 
in the calculations; essentially, this dependence is
similar to the geometrical parameters and the vibrational frequencies presented in 
% Secs.~\ref{sec-geometry} and \ref{sec-frequencies}. 
the Electronic Supplementary Information.
On the contrary, the adiabatic curves of the
radical anion exhibit a pronounced dependence on the basis set. 
A closer look at the $E_{A}(Q_f)$-curves reveals that it is not the 
size of the basis set that matters; of primary importance is to include diffuse basis functions.
For example, the results obtained by using the cc-pVTZ basis set are much poorer than 
by using the rather modest 6-31+g(d) basis set. Noteworthy is also the fact that 
an anion unstable against electron autodetachment (i.~e., $E_A > E_N$) can be the 
(incorrect) result of 
using an inadequate basis set; this aspect has already been pointed out.\cite{Kassab:96,Kassab:98}

%%%%%%%%%%%%%%%%%%%%%%%%%%%%%%%%%%%%%%%%%%%%%%%%%%%%%%%%%%%%%%%%%%%%%%%%%%%%%%%%%%%%%%%
%%%%%%%%%%%%%%%%%%%%%%%%%%%%%%%%%%%%%%%%%%%%%%%%%%%%%%%%%%%%%%%%%%%%%%%%%%%%%%%%%%%%%%
% We end this section by emphasizing the need to employ basis sets that include 
% diffuse functions in transport calculations, as an aspect of general importance for molecular electronics.
% The omission of the diffuse basis sets, even using 
% basis sets of double zeta quality (cc-pVDZ), as often the case 
% in many transport calculations, may yield errors $\sim 0.4 - 0.5$\,eV  in the LUMO energy offset. 
% Such errors are inacceptably large particularly in view of the fact that typical energy offsets
% amount $\sim 0.6 - 1.5$\,eV \cite{Venkataraman:12,Baldea:2012a,Baldea:2012b,Baldea:2012g}.
%%%%%%%%%%%%%%%%%%%%%%%%%%%%%%%%%%%%%%%%%%%%%%%%%%%%%%%%%%%%%%%%%%%%%%%%%%%%%%%%
\section{44BPY in solvents}
\label{sec-solvent}
%%%%%%%%%%%%%%%%%%%%%%%%%%%%%%%%%%%%%%%%%%%%%%%%%%%%%%%%%%%%%%%%%%%%%%%%%%%%%%%%
We have also extensively investigated the adiabatic energy surfaces of the neutral 
(44BPY$^0$) and charged (radical anion 44BPY$^{\bullet -}$ and radical cation 44BPY$^{\bullet +}$)
species in the solvents utilized in experiments, e.~g., 
aqueous solution,\cite{Tao:03,Tao:06b} toluene,\cite{Tao:05c} acetone,\cite{Venkataraman:12}
1,2,4-trichlorobenzene,\cite{Venkataraman:09} and acetonitrile.\cite{Kassab:96}
In line with the preliminary investigation of Ref.~\onlinecite{Baldea:2012i}, our calculations confirm the 
fact that, essentially, the solvent main effect is a nearly constant (i.~e., $\mathbf{Q}$-independent)
shift with respect to the adiabatic surfaces of the corresponding species in vacuo. 
Therefore, we do not present adiabatic surfaces in various solvents. Rather, we restrict ourselves 
to report results for
the aforementioned energy shift, which is important for molecular transport; 
the relative molecular orbital alignment 
with respect to the electrode Fermi levels is a key parameter that affects the current.\cite{Datta:03}

In Table \ref{table-solvent}, we present the computed influence of the solvents listed above
on the vertical ionization potentials (IP$=E_C - E_N = - E_{HOMO}$, where the subscript $C$ 
stands for radical cation) 
and the vertical electron affinities (EA$=E_N - E_A = - E_{LUMO}$) at the neutral equilibrium 
geometry. Although the 44BPY-based molecular junctions fabricated so far do not seem 
to exhibit hole-type (HOMO-mediated) conduction, we have also included 
results for the radical cation 44BPY$^{\bullet +}$ and for the HOMO 
in order to point out an aspect, which is
quite significant for the molecular transport in solvents. 

As visible in Table \ref{table-solvent}, the IP and EA-shifts are substantial and significantly dependent 
on the solvent. As a consequence, the HOMO-LUMO gap ${W}_{hl}$ is significantly affected by the solvents. 
Instead of estimating this gap from the HOMO and LUMO Kohn-Sham (KS) ``orbitals''
(whose physical meaning is questionable \cite{Gunnarson:89}), in Table \ref{table-solvent} 
we have determined it 
from $\Delta$-DFT calculations as ${W}_{hl} = $IP$-$EA$ =E_C + E_A - 2 E_N$, 
an expression which shows that it is the counterpart of what is called the charge gap 
in solid state and mesoscopic physics (see, e.~g., Refs.~\citenum{Baldea:2008,Baldea:2012i} and citations therein). 

The point to which we want to draw attention here is that, although the solvent substantially 
affects the HOMO-LUMO gap, it 
causes HOMO and LUMO shifts in opposite directions, and their magnitudes are practically 
\emph{equal}. The fact demonstrated here, that, for a given molecule (44BPY),
a variety of solvents cause opposite HOMO- and LUMO-displacements of almost same magnitudes 
is complementary to another fact known earlier; the HOMO and LUMO displacements were found 
to be similar for a variety of (smaller) molecules in a given (namely, aqueous) solution.\cite{Pearson:86}

%%%%%%%%%%%%%%%%%%%%%%%%%%%%%%%%%%%%%%%%%%%%%%%%%%%%%%%%%%%%%%%%%%%%%%%%%%%%%%%%%%%%%%%%%
%%%%%%%%%%%%%%%%%%%%%%%%%%%%%%%%%%%%%%%%%%%%%%%%%%%%%%%%%%%%%%%%%%%%%%%%%%%%%%%%%%%%%%%%%
%%%%%%%%%%%%%%%%%%%%%%%%%%%%%%%%%%%%%%%%%%%%%%%%%%%%%%%%%%%%%%%%%%%%%%%%%%%%%%%%%%%%%%%%%
\begin{table*} % [h!]
\small
\label{table-solution}
\begin{center}
% \begin{tabular*}{|l|@{\hspace{1ex}}c|@{\hspace{1ex}}c|@{\hspace{2ex}}c|@{\hspace{0ex}}c|@{\hspace{0ex}}c|@{\hspace{0ex}}c|@{\hspace{0ex}}c|@{\hspace{0ex}}}
\begin{tabular*}{\textwidth}{@{\extracolsep{\fill}}lllllll}
\hline
Medium         & ${W}_{hl}$      & IP$=-E_{HOMO}$ & EA$= - E_{LUMO}$   &  $\delta$\,IP   & $\delta$\,EA      &  $\delta$\,IP+$\delta$\,EA \\
\hline
vacuum         & 8.76; 8.69; 8.71        & 9.15; 9.13; 9.13     & 0.38; 0.44; 0.42        &   ---           &  ---              & ---                \\  
water          & 5.26; 5.21; 5.24        & 7.42; 7.41; 7.42     & 2.15; 2.21; 2.19        & -1.73; -1.72; -1.71    & 1.77; 1.76; 1.77        & 0.04; 0.04; 0.06         \\
toluene        & 6.73; 6.67; 6.68        & 8.13; 8.12; 8.12     & 1.40; 1.46; 1.44        & -1.01; -1.01; -1.01   & 1.02; 1.01; 1.01      & 0.001; 0.0005; 0.003     \\
acetone        & 5.40; 5.34; 5.37        & 7.48; 7.48; 7.49     & 2.08; 2.14; 2.12        & -1.66; -1.66; -1.64   & 1.70; 1.70; 1.70      & 0.04; 0.04; 0.05       \\
acetonitrile   & 5.32; 5.26; 5.29        & 7.44; 7.44; 7.45     & 2.12; 2.18; 2.16        & -1.70; -1.69; -1.68    & 1.74; 1.74; 1.74        & 0.04; 0.04; 0.06         \\
TCB            & 6.82; 6.75: 6.77        & 8.18; 8.17; 8.16     & 1.35; 1.41; 1.39        & -0.97; -0.97; -0.97   & 0.97; 0.97; 0.97      & 0.0001; -0.0006; 0.002    \\
\hline
\end{tabular*}
\caption{Values for the HOMO-LUMO gaps ${W}_{hl} \equiv E_{LUMO} - E_{HOMO}$, ionization potentials 
IP, and electron affinities of the 44BPY molecule in vacuo and several solvents (TCB$\equiv$\,1,2,4-trichlorobenzene) 
computed with the basis sets 6-31+g(d); aug-cc-pVDZ; aug-cc-pVTZ, respectively. 
Differences between the 
values of a certain quantity in solvent and in vacuo are denoted by $\delta$.
Notice the opposite signs and the almost 
equal magnitudes $\delta$\,IP\,$\simeq - \delta$\,EA
of the changes 
% $\delta$\,IP\,$\equiv\,$IP\,$-$\,IP$_{vac}=$ $-\delta$\,EA\,$\equiv-\,$EA\,$_{vac}+$\,EA
in the ionization potentials and electron affinities
caused by the solvents with respect to the values in vacuo.
See the main text for details.}
\label{table-solvent}
\end{center}
\end{table*}
%%%%%%%%%%%%%%%%%%%%%%%%%%%%%%%%%%%%%%%%%%%%%%%%%%%%%%%%%%%%%%%%%%%%%%%%%%%%%%%%%%%%%%
%%%%%%%%%%%%%%%%%%%%%%%%%%%%%%%%%%%%%%%%%%%%%%%%%%%%%%%%%%%%%%%%%%%%%%%%%%%%%%%%%%%%%%
\section{Impact of the twisted molecular conformation on the conductance}
\label{sec-G}
%%%%%%%%%%%%%%%%%%%%%%%%%%%%%%%%%%%%%%%%%%%%%%%%%%%%%%%%%%%%%%%%%%%%%%%%%%%%%%%%%%%%%%%
As already noted in Ref.~\onlinecite{Baldea:2012i} and reiterated above, 
unlike in the case of (practically) rigid molecules, 
the calculation of the electric current through molecular junctions based on floppy 
molecules requires a nontrivial step, namely an ensemble averaging 
in a nonequilibrium state.\cite{Baldea:2012i} % \cite{Baldea:2012i,Baldea:unpublished}. 
This ensemble averaging is a problem on its own that 
will not be addressed in this paper. % \cite{Baldea:unpublished}.

Instead, we will briefly consider a simpler problem related to it:
the dependence of the ohmic conductance $G$ on the twisting angle $\theta$.
Rather than examining changes in $\theta$ due to molecular vibrations,
in this section $\theta$ will be considered as an independent geometrical 
parameter. This problem is also of interest;
like in the case of the related biphenyl molecule,\cite{Venkataraman:06,Wandlowski:09}
one can synthesize and investigate, e.~g.,  
derivatives of a molecular 44BPY series wherein the torsion angle 
$\theta$ can be \emph{tuned} by 
bridging the two pyridyl rings with alkyl chains of \emph{variable} lengths.

The dependence $G = G(\theta)$, which has been considered theoretically 
for molecules consisting of two rings that can rotate relative to each 
other,\cite{Woitellier:89,Mujica:99,Haiss:06,Bagrets:08,Silva:10a,Reed:11,Pauly:12}
is interesting from the perspective of molecular conductance switching.
Theoretically, an approximate dependence $G \propto \cos^2 \theta$ has been found, 
which roughly agrees
with experimental data in biphenyl-derivatives.\cite{Venkataraman:06,Wandlowski:09}
This behavior characterizes a molecular switch that is ``on'' and ``off''
in parallel ($\theta = 0$) and perpendicular ($\theta = 90^\circ$) conformations,
respectively. 

\ib{
In spite of significant differences between 
existing approaches of the (ohmic) transport in 44BPY,
\cite{Hou:05,Thygesen:05c,Bagrets:08,Venkataraman:12}
which are based on nonequilibrium Green's function (NEGF) methods and 
the Landauer trace formula (see, e.~g., Refs.~\citenum{Datta:05,Pati:08,Baldea:2010f}),
there is an important common finding in those studies. Namely, 
the fact that the coherent electron transport is determined by the LUMO: 
close to the metallic Fermi energy $E_F$, the calculated 
transmission exhibits a pronounced Lorentzian peak, which is well separated 
energetically from other peaks situated away from $E_F$.
This single-level (``only LUMO'') model is simple but turned out to be 
realistic for 44BPY. By interpreting experimental 
thermopower data within this model,
the LUMO was found to lie by $\sim 1.53$\,eV above the metallic Fermi energy,
in good agreement with the theoretically estimated value of 
$\sim 1.47$\,eV.\cite{Venkataraman:12}}
So, we have a strong reason to accept this model, and then 
the low-bias conductance $G$ can be expressed as
(see, e.~g., Refs.~\onlinecite{Datta:05,Baldea:2012a})
%%%%%%%%%%%%%%%%%%%%%%%%%%%%%%%%%%%%%%%%%%%%%%%%%%%%%%%%%%%%%%%%%%%%%%%%%%%%%%%%%%%%%%%
\begin{equation}
\label{eq-G}
g(\theta) \equiv \frac{G(\theta)}{G_0} 
= \frac{\Gamma^2}{\tilde{E}_{LUMO}^{2}(\theta) + \Gamma^2} 
\simeq \frac{\Gamma^2}{\tilde{E}_{LUMO}^{2}(\theta)} .
\end{equation}
%%%%%%%%%%%%%%%%%%%%%%%%%%%%%%%%%%%%%%%%%%%%%%%%%%%%%%%%%%%%%%%%%%%%%%%%%%%%%%%%%%%%%%% 
Here $\Gamma$ is a finite width due to the LUMO-electrode couplings,\cite{Pati:06}
which is usually much smaller than the LUMO-energy offset $\tilde{E}_{LUMO}$ 
relative to the Fermi energy $E_F$ of the electrodes to which the molecule is linked.
The quantity $\tilde{E}_{LUMO}$ differs from the difference between the value 
$E_{LUMO}$ of the isolated molecule and $E_F$, because of the energy shift 
$\delta E$ caused by molecule-electrode interactions 
(e.~g., image charge effects \cite{desjonqueres:96}). We do not attempt to 
microscopically estimate
here the difference between $\tilde{E}_{LUMO}$ and $E_{LUMO}$
($\tilde{E}_{LUMO} = E_{LUMO} - E_F - \delta E$). We determine this shift by 
requiring that the value of $\tilde{E}_{LUMO}(\theta_{eq})$ at the equilibrium value
$\theta = \theta_{eq}$ (cf.~Table S1) % Table \ref{table-neutral}) 
coincides with that 
deduced experimentally $\tilde{E}_{LUMO}(\theta_{eq}) \simeq 1.5$\,eV for 44BPY 
junctions.\cite{Venkataraman:12}
Based on the previous calculations,\cite{Baldea:2012i} which indicated that
the metal (Au) atoms contacted to 44BPY essentially causes a constant ($Q_f$-independent) 
shift of the LUMO energy (compare Figs.~3a and 3c of Ref.~\onlinecite{Baldea:2012i}), we will
ignore the $\theta$-dependence of $\tilde{E}_{LUMO}(\theta) - E_{LUMO}(\theta)$.
For the basis set aug-ccpVTZ, we thus deduced $\tilde{E}_{LUMO}(\theta) - E_{LUMO}(\theta) \simeq 1.92$\,eV.

Further, 
in view of the fact that changes of the molecule-electrode contacts turn out to only 
weakly affect
the torsional angle,\cite{Ratner:09} we will also neglect a possible $\theta$-dependence of 
$\Gamma$; we do not expect a substantial change in the short-range (\emph{contact}) interaction 
between the LUMO and electrodes by varying the \emph{inter}-ring angle $\theta$.
By employing the experimental values 
$G(\theta_{eq})/G_0 = 0.00068$ and $\tilde{E}_{LUMO}(\theta_{eq}) \simeq 1.5$\,eV,\cite{Venkataraman:12}
Eq.~(\ref{eq-G}) yields the value $\Gamma \simeq 0.039$\,eV.

\begin{figure}[h!]
$ $\\[6ex]
% 
% \centerline{\hspace*{-0ex}\includegraphics[width=0.4\textwidth,angle=0]{../g/fig_phi_vac_ccpvdz_ccpvtz_augccpvdz_augccpvtz_631+g_d.eps}}
\centerline{\hspace*{-0ex}\includegraphics[width=0.4\textwidth,angle=0]{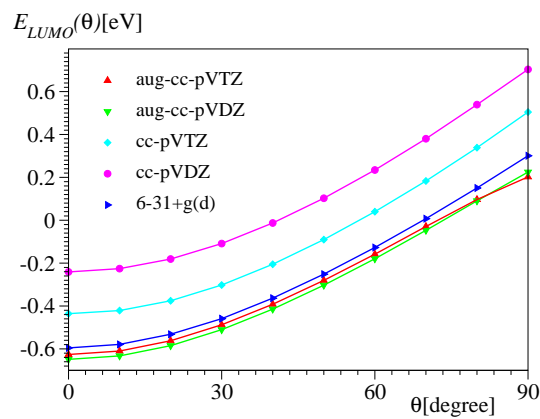}}
$ $\\[6ex]
\caption{$\Delta$-DFT results for the LUMO energy of the 44BPY molecule 
computed with several basis sets (specified in the inset). 
Notice the importance of the diffuse functions. See the main text for details.}
\label{fig:e-lumo-theta}
\end{figure}

To obtain the LUMO energy $E_{LUMO}(\theta)$ of the isolated molecule, 
we performed $\Delta$-DFT-calculations \cite{Gunnarson:89}
and computed \cite{not-FC} 
%%%%%%%%%%%%%%%%%%%%%%%%%%%%%%%%%%%%%%%%%%%%%%%%%%%%%%%%%%%%%%%%%%%%%%%%%%%%%%%%%%%%%%%
\begin{equation}
\label{eq-delta-dft}
E_{LUMO}(\theta) = E_A (\theta; \mathbf{R^\prime})- E_N (\theta; \mathbf{R^\prime}) .
\end{equation}
%%%%%%%%%%%%%%%%%%%%%%%%%%%%%%%%%%%%%%%%%%%%%%%%%%%%%%%%%%%%%%%%%%%%%%%%%%%%%%%%%%%%%%%
For simplicity, we present results of the calculations carried out for various conformations 
by varying the torsion angle ($\theta$) while freezing all the other geometrical 
parameters at the neutral equilibrium geometry (denoted by $\mathbf{R^\prime}$).\cite{theta-dependence}
Results illustrating the $\theta$-dependence of the LUMO energy $E_{LUMO}(\theta)$ of 
the isolated molecule are shown in Fig.~\ref{fig:e-lumo-theta}.

%%%%%%%%%%%%%%%%%%%%%%%%%%%%%%%%%%%%%%%%%%%%%%%%%%%%%%%%%%%%%%%%%%%%%%%%%%%%%%%%%%%%%% 
% \begin{figure}[h!]
% $ $\\[6ex]
% \centerline{\hspace*{-0ex}\includegraphics[width=0.4\textwidth,angle=0]{../g/fig_phi_vac_pcm_water_toluene_124TCB_acetone_augccpvdz.eps}}
% $ $\\[6ex]
% \caption{$\Delta$-DFT (B3LYP + aug-cc-pVDZ) results (points) for the LUMO energy of the 44BPY molecule in vacuo and several solvents.
% The lines represent results of fitting by usings Eq.~(\ref{eq-fit}). See the main text for details.}
% \label{fig:phi-vac-various-pcm}
% \end{figure}
%%%%%%%%%%%%%%%%%%%%%%%%%%%%%%%%%%%%%%%%%%%%%%%%%%%%%%%%%%%%%%%%%%%%%%%%%%%%%%%%%%%%%%
%%%%%%%%%%%%%%%%%%%%%%%%%%%%%%%%%%%%%%%%%%%%%%%%%%%%%%%%%%%%%%%%%%%%%%%%%%%%%%%%%%%%%%%
%%%%%%%%%%%%%%%%%%%%%%%%%%%%%%%%%%%%%%%%%%%%%%%%%%%%%%%%%%%%%%%%%%%%%%%%%%%%%%%%%%%%%%%

\begin{figure}[h!]
$ $\\[6ex]
% 
% \centerline{\hspace*{-0ex}\includegraphics[width=0.4\textwidth,angle=0]{../g/fig_g_vs_cos_Sq_phi.eps}}
\centerline{\hspace*{-0ex}\includegraphics[width=0.4\textwidth,angle=0]{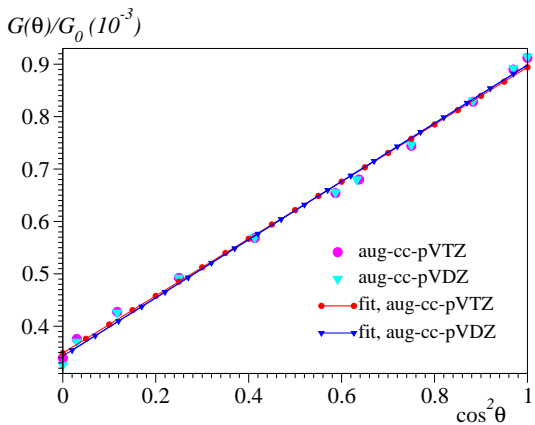}}
%% \centerline{\hspace*{-0ex}\includegraphics[width=0.4\textwidth,angle=0]{../g/fig_gcalc_gfit_augccpvdz_augccpvtz_vs_phi.eps}}
%% \centerline{\hspace*{-0ex}\includegraphics[width=0.4\textwidth,angle=0]{../g/fig_phi_vac_augccpvdz_b3lyp_bp86.eps}}
$ $\\[6ex]
\caption{Dependence of the conductance $G$ in units of conductance quantum $G_0=2 e^2/h = 77.5\,\mu$S 
on the torsional angle $\theta$ obtained by using the basis sets specified in the inset.
For each basis set, the conductance has been been fixed at the experimental value $G(\theta_{eq})/G_0 = 0.00068$
\cite{Venkataraman:12} ($\theta_{eq} = 37.01$ for aug-cc-pVDZ and $\theta_{eq} = 37.3$ for aug-cc-pVTZ, 
cf.~Table S1). % Table \ref{table-neutral}).
The straight lines have been obtained by fitting the results of quantum chemical calculations using 
the Ansatz of Eq.~(\ref{eq-G-ansatz}).
\ib{The fact that the symbols corresponding to the various basis sets employed are hardly distinguishable
from each other demonstrates the realiability of the results.}
See the main text for details.}
\label{fig:g-vs-cosSq}
\end{figure}

Once  $\tilde{E}_{LUMO}(\theta)$ is known, 
the $\theta$-dependence of the conductance can be deduced by means of 
Eq.~(\ref{eq-G}). The dependence of $G$ on $\cos ^2 \theta$ is shown in 
Fig.~\ref{fig:g-vs-cosSq}. As visible there, our result does not give support 
to the proportionality $G \propto \cos ^2 \theta$ found 
experimentally for other molecules \cite{Venkataraman:06,Wandlowski:09} 
and DFT-based theories approaches.\cite{Bagrets:08,Silva:10a,Pauly:12}
Our results agree very well with the following Ansatz,
which straightforward generalizes the aforementioned relationship
%%%%%%%%%%%%%%%%%%%%%%%%%%%%%%%%%%%%%%%%%%%%%%%%%%%%%%%%%%%%%%%%%%%%%%%%%%%%%%%%%%%%%%%
\begin{equation}
\label{eq-G-ansatz}
G(\theta)/G_0 = g_{\perp} + \left(  g_{\parallel} -  g_{\perp}\right) \cos^2 \theta .
\end{equation}
%%%%%%%%%%%%%%%%%%%%%%%%%%%%%%%%%%%%%%%%%%%%%%%%%%%%%%%%%%%%%%%%%%%%%%%%%%%%%%%%%%%%%%% 

It is worth emphasizing at this point the fact that in 44BPY the 
relationship $G \propto \cos ^2 \theta$, which yields 
a perfect ``off'' switch $G(\theta = 90^\circ) \equiv 0$, 
is contradicted by a detailed microscopical analysis.\cite{Woitellier:89} 
According to Ref.~\onlinecite{Woitellier:89}, 
the transmission through 44BPY remains nonvanishing 
even in a strictly perpendicular conformation, due to the existence of crossed 
$\sigma - \pi$ electronic interaction.\cite{Woitellier:89} According to the present 
estimation, $g_{\perp} \simeq g_{\parallel}/3$.
%%%%%%%%%%%%%%%%%%%%%%%%%%%%%%%%%%%%%%%%%%%%%%%%%%%%%%%%%%%%%%%%%%%%%%%%%%%%%%%%%%%%%%%
%%%%%%%%%%%%%%%%%%%%%%%%%%%%%%%%%%%%%%%%%%%%%%%%%%%%%%%%%%%%%%%%%%%%%%%%%%%%%%%%%%%%%%%

\ib{We end this sections with two remarks. First, we emphasize 
the difference between our present $\Delta$-DFT-based calculations 
and most DFT-approaches, wherein mathematical objects, namely KS orbitals,
are treated as if they were real physical orbitals. Second, we note the relationship 
of the $\theta$-dependence discussed above and the 
very small energy barrier associated to the profile of the torsional potential, which
is $E_N(\theta)$ in our notation. An important issue amply discussed in the past 
(see, e.~g., Refs.~\citenum{Kassab:98,Perez:05b} and citations therein)
is the magnitude of the barrier heights between the twisted global minimum
and the planar and perpendicular conformers, 
$\Delta E_0 \equiv E_{N}(\theta = 0^\circ) - E_{N}(\theta_{eq})$ 
and 
$\Delta E_{90} \equiv E_{N}(\theta = 90^\circ) - E_{N}(\theta_{eq})$,
respectively.
The most recent calculations published so far \cite{Perez:05b} succeded to
produce values $\Delta E_0 \simeq 1.5$\,kcal/mol and $\Delta E_{90} \simeq 2.2$\,kcal/mol, 
which agree with the ordering $\Delta E_0 <  \Delta_{90}$ deduced
from measured NMR spectra of 44BPY dissolved in nematic liquid crystal.\cite{Emsley:75}
However, there still remain significant differences between 
theory and experiment; proton NMR measurements estimated 
$\Delta E_0 \simeq 4.0$\,kcal/mol.\cite{Mangutova:73}
Internal rotational barriers of 
heights comparable to 44BPY have also been found 
for other molecular systems.\cite{Tsuzuki:99,AyanDatta:10}
}
%%%%%%%%%%%%%%%%%%%%%%%%%%%%%%%%%%%%%%%%%%%%%%%%%%%%%%%%%%%%%%%%%%%%%%%%%%%%%%%%
\section{Discussion and Conclusions}
\label{sec-conclusion}
%%%%%%%%%%%%%%%%%%%%%%%%%%%%%%%%%%%%%%%%%%%%%%%%%%%%%%%%%%%%%%%%%%%%%%%%%%%%%%%%
In this paper, we have reported results of extensive quantum chemical calculations for the 
44BPY molecule in vacuo and in different solvents employed in relevant experiments.

The small differences between the measured properties of the neutral 
44BPY$^0$ and radical anion 44BPY$^{\bullet -}$
and those, nearly convergent with respect to the basis set size, calculated 
within the DFT/B3LYP presented in the Electronic Supplementary Information
% in Secs.~\ref{sec-geometry} and \ref{sec-frequencies}
are representative for the present state-of-the-art of the theory:
on one side, it demonstrate its rather high accuracy and, on the other side,  
that intrinsic limitations remain.

The results on the adiabatic energy curves and on the solvent effects 
presented in this work
% Secs.~\ref{sec-adiabatic-curves} and \ref{sec-solvent}
can (and will) 
serve as input information for subsequent transport studies through 44BPY-based molecular junctions.

Because, as expressed by the title, the remote goal of the present 
investigation is the molecular transport, three findings of this work are 
particularly worth to emphasize. 

(i) We are not aware of attempts to microscopically incorporate solvent effects into transport 
approaches. 
Available program packages do not allow to perform microscopic transport 
calculations through molecular devices immersed in electrolytes. 
Our results indicate that the solvent acts as a \emph{selective} gate electrode:
it causes energy shifts of particle-like and hole-like orbitals
% the HOMO (an occupied orbital) and the LUMO (an unoccupied orbital) 
in \emph{opposite} directions and nearly same magnitudes.
On this basis, we propose to embody solvent effects 
on molecular transport within a procedure based on a scissor operator.
Let us remind at this point that the scissor operator technique has been proposed 
\cite{Schlueter:84} and applied 
\cite{Godby:88,Levine:89,Baldereschi:89,Baldereschi:95,Gonze:95,Gonze:97b}
in semiconductors to
empirically cut the band structure along the band gap and rigidly move 
the energies of the conduction and valence bands in opposite directions. 
There, the scissor operator has been motivated technically, as 
an (empirical) correction of the band gap, whose value is underestimated by the DFT.
More recently, a scissor operator has been utilized in molecular electronics 
\cite{Neaton:06,Quek:07,Venkataraman:12} with both a technical and a physical motivation: 
to correct the too narrow KS HOMO-LUMO gap
and to account for the energy shift due to electrodes via image effects,\cite{desjonqueres:96}
respectively. For simplicity, in those approaches 
the HOMO and all occupied orbitals,
and the LUMO and all unoccupied orbitals are shifted in opposite directions by the same quantity $\Delta/2$
(half the correction to the HOMO-LUMO gap).

Similar to those discussed above, we also propose to account for solvent effects on the molecular transport 
by means of a scissor operator: the energies of the occupied and unoccupied 
orbitals should be moved by $-\delta$\,IP and by $-\delta$\,EA, respectively.
Unlike in the aforementioned cases, wherein the assumption that the opposite shifts have the 
\emph{same} magnitude ($\Delta/2$) 
is merely done for convenience, our results give a microscopic support to the fact
that the shifts are opposite and have  
practically the \emph{same} magnitude
$-\delta$\,IP$\simeq + \delta$\,EA. Notice that (a) $\delta$\,IP and $\delta$\,EA
are not empirical quantities but deduced
from quantum chemical calculations, and (b) the presently proposed scissor operator procedure 
can be applied on top of any transport approach in vacuo, let it be a `DFT+$\Sigma$'' implementation
\cite{Neaton:06,Quek:07,Venkataraman:12} or else.

(ii) The results presented here % in Secs.~\ref{sec-adiabatic-curves} and \ref{sec-G}
(Figs.~\ref{fig:e-Q-f-basis} and \ref{fig:e-lumo-theta})
clearly demonstrate that, in order to correctly describe the anion (thence the LUMO),
it is necessary to employ basis sets that include diffuse basis functions. 
Even by using basis sets of double zeta quality (cc-pVDZ), 
the omission of the diffuse functions, 
as often the case in many transport calculations, 
may yield errors $\sim 0.4 - 0.5$\,eV  in the LUMO energy offset. 
In view of the fact that typical energy offsets
amount $\sim 0.6 - 1.5$\,eV,\cite{Venkataraman:12,Baldea:2012a,Baldea:2012b,Baldea:2012g}
such errors in determining the LUMO location are unacceptably large.

(iii) Along with molecular junctions wherein a single 44BPY molecule 
is directly linked to electrodes,\cite{Tao:03,Tao:05c,Tao:06b,Venkataraman:12}
molecular junctions wherein spacers (e.~g., six alkyl units at either side
\cite{Wandlowski:08}) are used to link the 44BPY molecule to electrodes 
are also of experimental interest. Because the largest basis sets employed 
in the present work may become a challenge for such molecular sizes,
it is important to note that, once diffuse basis functions are included, 
anionic species can still be reliably described with rather modest basis sets,
as illustrated by the results obtained by employing the 6-31+g(d) basis set
shown in Figs.~\ref{fig:e-Q-f-basis} and \ref{fig:e-lumo-theta} and
Table~\ref{table-solvent}.

Figs.~\ref{fig:e-Q-f-basis} and \ref{fig:e-lumo-theta} present results 
demonstrating both the importance of (a) the diffuse basis functions 
and of (b) the most salient structural feature of the 44BPY molecule,
namely the twisting inter-ring angle.
One should emphasize that these two aspects are distinct from each other.
Concerning aspect (a), one should note that 
diffuse functions are needed for the correct description of the anion. 
The errors introduced by using basis sets without diffuse functions,
e.~g., at the neutral equilibrium geometry 
($Q_f = 0$ in Fig.~\ref{fig:e-Q-f-basis} or $\theta =  \theta_{eq} \simeq 37^\circ$
in Fig.~\ref{fig:e-lumo-theta}) are practically the same as those far 
away from it (cf.~Figs.~\ref{fig:e-Q-f-basis} and \ref{fig:e-lumo-theta}).
Concerning aspect (b), we remind what was already stated 
in Ref.~\onlinecite{Baldea:2012i} (see especially the next-to-last section of that work): 
the floppy degree of freedom ($Q_f$)
as well as its strong anharmonicity and significant reorganization energy are
properties of the 44BPY molecule, which are affected neither by solvents 
nor by electrodes. Computations for 44BPY$^{0}$ and 44BPY$^{\bullet -}$ 
immersed in solvents and by attaching gold atoms at the two ends show that the main 
effect is a practically constant shift of the anionic adiabatic energy surface with 
respect to the anion placed in vacuo.

To account for the effect of the intramolecular reorganization related to the floppy 
degree of freedom on the electron transport through 44BPY-based junctions, 
one should perform an ensemble averaging on a molecular system at away from equilibrium 
(nonvanishing external bias), which is a nontrivial problem.\cite{Baldea:2012i} 
As a preliminary step in demonstrating the significant impact of 
molecular conformation on the transport, % in Sec.~\ref{sec-G}
we have presented results for the linear (ohmic) conductance
as a function of the torsional angle $G=G(\theta)$. 
Our results obtained by quantum chemical calculations
are well described by a phenomenological Ansatz, Eq.~(\ref{eq-G-ansatz}).
% This result differs from most of the previous results 
% value found for the conductance in a perpendicular conformation, 
% which is significantly different from zero ($G_{\perp} \simeq G_{\parallel}/3$).

To end, let us finally mention that, 
at $\theta = 90^\circ$, the 44BPY molecule 
(D$_{2d}$ symmetry) becomes unstable against a Jahn-Teller distortion.\cite{Lauher:80,Woitellier:89}
Therefore, significant deviations from the above Ansatz may occur 
close to the perpendicular conformation. 
Obviously, this is an issue that escapes the present considerations.
%%%%%%%%%%%%%%%%%%%%%%%%%%%%%%%%%%%%%%%%%%%%%%%%%%%%%%%%%%%%%%%%%%%%%%%%%%%%%%%%
\subsubsection*{Acknowledgments~~}
Financial support from
the Deu\-tsche For\-schungs\-ge\-mein\-schaft % (DFG) 
is gratefully acknowledged.
% 
% \balance
%%%%%%%%%%%%%%%%%%%%%%%%%%%%%%%%%%%%%%%%%%%%%%%%%%%%%%%%%%%%%%%%%%%%%%%%%%%%%%%%
\renewcommand\refname{Notes and references}
\footnotesize{
%%Rev \bibliographystyle{rsc}
%%Rev \bibliography{/home/ioan/QDs/LinearResponse/paper/bibl,/home/ioan/QDs/LinearResponse/paper/misc}

\begin{mcitethebibliography}{77}
\bibitem[not()]{notationPlusAst}
The notation ``(+*)'' means that diffuse and polarization functions have been
  added only to the nitrogen atoms.\relax
\mciteBstWouldAddEndPunctfalse
\mciteSetBstMidEndSepPunct{\mcitedefaultmidpunct}
{}{\mcitedefaultseppunct}\relax
\EndOfBibitem
\bibitem[Ann()]{Annika}
Thanks are due to Annika Bande for help to calculate the bond orders.\relax
\mciteBstWouldAddEndPunctfalse
\mciteSetBstMidEndSepPunct{\mcitedefaultmidpunct}
{}{\mcitedefaultseppunct}\relax
\EndOfBibitem
\end{mcitethebibliography}
%%REv \end{document}
%%%%%%%%%%%%%%%%%
\providecommand*{\mcitethebibliography}{\thebibliography}
\csname @ifundefined\endcsname{endmcitethebibliography}
{\let\endmcitethebibliography\endthebibliography}{}

}
\balance
\end{document}